\documentclass[useAMS,usenatbib]{mn2e}

\usepackage[dvips]{color,graphicx}

\usepackage{amsmath}

\newcommand{\del}[2]%
{\frac{\mathrm{d}{#1}}{\mathrm{d}{#2}}}
\newcommand{\Del}[2]%
{\frac{\mathrm{D}{#1}}{\mathrm{D}{#2}}}
\newcommand{\ddel}[2]%
{\frac{\mathrm{d}^2{#1}}{\mathrm{d}{#2}^2}}

\newcommand{\pdel}[2]%
{\frac{\partial{#1}}{\partial{#2}}}
\newcommand{\pddel}[2]%
{\frac{\partial^2{#1}}{\partial{#2}^2}}

\newcommand{\simgt}{\lower.5ex\hbox{$\; \buildrel > \over \sim \;$}}
\newcommand{\simlt}{\lower.5ex\hbox{$\; \buildrel < \over \sim \;$}}

\newcommand{\bracket}[1]{\big<#1\big>}

\newcommand{\bu}{{\bf u}}
\newcommand{\vep}{\varepsilon}

\def\e{\mathrm{e}}

\def\n{\mathrm{n}}
\def\p{\mathrm{p}}
\def\u{\mathrm{u}}

\def\Rnum#1{\uppercase\expandafter{\romannumeral #1}}
\def\rnum#1{\expandafter{\romannumeral #1}}

\title[Neutrino Acceleration by Bulk Matter Motion]
{Neutrino Acceleration by Bulk Matter Motion and Explosion Mechanism of Gamma-Ray Bursts}
\author[Yudai Suwa]{Yudai Suwa$^{1}$\thanks{E-mail: suwa@yukawa.kyoto-u.ac.jp} 
\\
$^{1}$Yukawa Institute for Theoretical Physics, Kyoto University,
Oiwake-cho, Kitashirakawa, Sakyo-ku, Kyoto, 606-8502, Japan}

\begin{document}

\date{draft version on \today}

\pagerange{\pageref{firstpage}--\pageref{lastpage}} \pubyear{2012}

\maketitle

\label{firstpage}

\begin{abstract}
The neutrino annihilation is one of the most promising candidates for
the jet production process of gamma-ray bursts. Although neutrino
interaction rates depend strongly on the neutrino spectrum, the
estimations of annihilation rate have been done with an assumption of
the neutrino thermal spectrum based on the presence of the
neutrinospheres, in which neutrinos and matter couple strongly. We
consider the spectral change of neutrinos caused by the scattering by
infalling materials and amplification of the annihilation rate. We
solve the kinetic equation of neutrinos in spherically symmetric
background flow and find that neutrinos are successfully accelerated
and partly form nonthermal spectrum. We find that the accelerated
neutrinos can significantly enhance the annihilation rate by a factor
of $\sim 10$, depending on the injection optical depth.
\end{abstract}

\begin{keywords}
acceleration of particles -- accretion, accretion discs -- gamma-ray
burst: general -- neutrinos -- radiative transfer
\end{keywords}

\section{Introduction}

The neutrinos play a very important role in the extremal condition in
astrophysics.  Although photons are strongly coupled with the matter
in the dense material, neutrinos are able to escape due to its weak
coupling with the matter.  Therefore, neutrinos can be an important
cooling source and significantly affect the dynamics.  In addition,
they can even be a heating source in several cases, e.g., the neutrino
capture process in the core-collapse supernovae (CCSNe) \citep{beth90}
and the neutrino annihilation in the central engine of gamma-ray
bursts (GRBs) \citep{macf99}.  In the hot (temperature $T\simgt 1$MeV)
and dense (density $\rho\simgt 10^{11}$ g cm$^{-3}$) gas, even
neutrinos are trapped and thermalized so that the temperature of
neutrinos becomes coincident with that of the matter. The surface
where the neutrinos are decoupled from the matter is called
``neutrinosphere'', which is analogous to a photosphere of photons.

Due to the existence of neutrinosphere, the spectrum of neutrinos is
often assumed to be a thermal (Fermi-Dirac) distribution. However, we
know, by the numerous studies of photons, that the radiation spectrum
can easily be deformed from the thermal one in the propagation
regime. The nonthermal radiation spectrum can be produced by the
nonthermal spectrum or the different temperature of the scattering
bodies.
As for the case of photons, the processes that produce the nonthermal
spectrum by electron scattering are called thermal Compton \citep[see,
  e.g.,][]{rybi79} and bulk Compton processes \citep{blan81}. The
former process is driven by the electrons with different temperature
from photons, while the latter is induced by electrons with the
inhomogeneous velocity field. In the case of neutrinos, the bulk
motion of scattering materials could lead to the similar effect and
produce nonthermal component of neutrinos, which was not investigated
so far.\footnote{Indeed, the term which is related to this effect is
  included in the numerical simulations that solve the neutrino
  Boltzmann equation.  However, due to the small radial velocity in
  the postshock region, this effect plays significantly minor role in
  the context of the core-collapse supernovae. Thus, there was no
  study focusing on this effect. On the other hand, we consider the
  free-fall background flow without the shock in this paper so that
  the radial velocity is large enough to make nonthermal component. }

Neutrino interactions with matter strongly depend on the energy of
neutrinos, i.e., the cross section $\sigma\propto \vep_\nu^2$ with
$\vep_\nu$ being the neutrino energy. Thus, little difference of
spectrum (especially at the high-energy region) could lead
significantly different dynamics. The neutrino capture and the
neutrino annihilation are critically important for the shock revival
of CCSNe and the jet production of GRBs,\footnote{Note that there are
  other alternatives for jet production mechanism of GRBs. Among them,
  another promising candidate is the Poynting dominated jet, which is
  driven by the magnetic field.} respectively. As for CCSNe, there are
significant efforts for solving Boltzmann equation of neutrinos with
hydrodynamics numerically because the explosion mechanism
(particularly delayed-explosion scenario) tremendously relies on
neutrino physics. Since the neutrinospheres for CCSNe are basically
spherical (deformation is moderate if the rotation is not very rapid),
the neutrino transfer can be solved with spherical symmetric
background, which is reachable even for the current computer
resources. In fact, hydrodynamic simulations together with neutrino
Boltzmann equation have been done in spherical symmetric case (e.g.,
\citealt{lieb01,sumi05}).\footnote{Recently, multi-dimensional
  hydrodynamic simulations with neutrino transfer have been performed
  by several groups \citep{bura06a,burr06,brue09,suwa10} by employing
  several assumptions to reduce the computational costs. More
  recently, the development of a full seven-dimensional Boltzmann
  solver is reported \citep{sumi12}.} On the other hand, the central
engine of GRBs is essentially asymmetric (e.g., the compact object and
the accretion disk system) so that neutrino transfer should be solved
with multi-dimensional treatment. Although there are a few attempts
to solve the neutrino radiative transfer in multidimensional manner
\citep[e.g.,][]{dess09,sumi12}, the long-term dynamical simulation is
still too computationally expensive so that the numerical solutions of
full Boltzmann equation are not accessible at the moment.  Because of
these facts, the neutrino interactions in the central engine of GRBs
are introduced with plenty of assumptions. One of them is the thermal
spectrum.

Among the accretion disk models, the neutrino-dominated accretion flow
(NDAF), in which copious neutrinos are emitted and dominates the
cooling, is often discussed as a candidate of the central engine of
GRBs.  \cite{poph99} derived the disk structure and neutrino
luminosity by solving the group of equations of state, hydrodynamics,
thermodynamics and microphysics in detail.  The energy conversion
efficiencies become extremely high (the annihilation luminosity
$L_{\nu\bar\nu}$ becomes as large as $\sim 10^{53}$ erg s$^{-1}$) for
the mass accretion rate $\dot M$ = 10$M_\odot$ s$^{-1}$. However,
their results are too optimistic, as they ignored neutrino opacity and
overestimated the neutrino luminosity.  \cite{dima02} showed that the
effect of neutrino opacity becomes significant for $\dot M > 1M_\odot$
s$^{-1}$, and recalculated the annihilation rate including the concept
of neutrinosphere. They demonstrated that the $L_{\nu\bar\nu}$
increases up to its maximum value of $\sim 10^{50}$ erg s$^{-1}$ at
$\dot M \approx 1M_\odot$ s$^{-1}$ and decrease for larger $\dot
M$. Thus they concluded that the neutrino annihilation in NDAF is not
a sufficient mechanism for liberating large amount of energy.
\cite{naga07} performed axisymmetric simulation of the collapsar and
found that the neutrino annihilation is less important than neutrino
capture as a heating source.  These negative results could come from
the assumption employed in their calculation, i.e., they neglected the
neutrino emission from the region with neutrino optical depth
$\tau\simgt 1$ and employed a thermal distribution with a single
temperature for the neutrino spectrum.\footnote{ There are a large
  number of attempts to amplify the neutrino-annihilation
  rate. Especially, the effects of (general) relativity are paid
  attention, such as beaming by relativistic motion and bending by
  black-hole space time \citep[e.g.,][]{asan00,asan01}. The
  relativistic effects for the structure of NDAF is also discussed
  recently \citep{chen07,zala11}.}

In this paper, multiple scattering of neutrinos and the acceleration
(i.e., up scattering) in a fluid flow is considered. We investigate
the impact to the pair-annihilation rate by accelerated component of
neutrinos. Note that although in this paper we consider the parameter
regime for the collapsar scenario that is one of promising candidates
of long-duration GRBs, the neutrino acceleration process, however, is
viable for the central engine of short GRBs. Thus, the following is
applicable both for long and short GRBs.  The rest of this paper is
organized as follows. In \S \ref{sec:neutrinosphere}, we briefly
review the concept of neutrinosphere. In \S \ref{sec:radiative}, we
describe the radiative transfer equation for neutrinos and show that
its solution contains nonthermal component. In \S
\ref{sec:annihilation} we investigate the effect of nonthermal
component of neutrino on the neutrino-annihilation rate.  We summarize
our results and discuss their implications in \S \ref{sec:summary}.

\section{Neutrinospheres}
\label{sec:neutrinosphere}

There are three types of neutrinosphere, which are determined by the
different micro processes \citep[see][]{raff01}. Here, we explain
these ``neutrinospheres'' one by one:
\begin{itemize}
\item {\bf Number sphere}: The optical depth by the emission and
  absorption of neutrinos is about unity. As for $\nu_\e$ and
  $\bar\nu_\e$, the electron/positron capture and its inverse process
  (i.e., $\nu_\e + \n \leftrightarrow \p+\e^-$ and $\bar\nu_\e + \p
  \leftrightarrow \n+\e^+$) are important processes. As for $\nu_X$,
  which represents heavier leptonic neutrinos and their antineutrinos
  (i.e., $\nu_\mu, \nu_\tau,\bar\nu_\mu,$ and $\bar\nu_\tau$), the
  pair production/annihilation processes (i.e.,
  $\nu\bar\nu\leftrightarrow\gamma\gamma$, $\nu\bar\nu\leftrightarrow
  e^+e^-$, $NN\leftrightarrow NN\nu\bar\nu$) determine the opacity.
\item {\bf Energy sphere}: The inelastic scattering by electron is
  important process here. The electrons receive the energy from
  neutrinos because the electron rest mass energy (511 keV) is much
  smaller than the typical neutrino energy ($\sim 10$ MeV), which is
  determined by the matter temperature at the number sphere. Inside
  the energy sphere the neutrinos are thermalized due to energy
  transfer with electrons, which are tightly coupled with baryons.
\item {\bf Transport sphere}: Beyond the energy sphere, the elastic
  scattering by nucleons and nuclei is dominant source of the
  opacity. Because the rest mass energy of these particles is much
  larger than neutrino energy, these scattering can be treated as the
  elastic scattering.\footnote{Note that \cite{raff01} investigated
    how the recoil term affects the spectrum of $\nu_X$.}
\end{itemize}

As for $\nu_\e$ and $\bar\nu_\e$, all neutrinospheres provided above
are almost coincident so that the spectrum is almost thermal. On the
other hand, $\nu_X$ has distinct radii of neutrinospheres
\citep{raff01}. Therefore, $\nu_X$ could have nonthermal component,
which would be produced between energy and transport spheres.

\section{The Radiative Transfer Equation and Its Solution}\label{sec:radiative}

Here we consider the neutrino radiative transfer on the background of
hyperaccreting matter. We focus on the region between the energy
sphere and the transport sphere, where the scattering by nucleons
dominates the opacity.

\subsection{Accretion Flow}

Here, we briefly describe profiles of matter as a background of
neutrino radiative transfer. In this calculation, we employ the
spherically symmetric accretion flow in order to mimic the collapsar.

Let $\dot M$ be the rate at which matter is accreting, and let its
radial inward speed be
\begin{equation}
  u(r)=c\left(\frac{r_s}{r}\right)^{1/2},
  \label{eq:ur}
\end{equation}
where $c$ is the speed of light and $r_s$ is the Schwarzschild
radius. This free-fall velocity profile makes the implicit assumption
that the radiation force on the accreting matter is insignificant or
equivalently, that the escaping luminosity is much less than the
Eddington value. The scattering optical depth of the flow from a
radius $r$ to infinity is given by
\begin{equation}
\tau_\mathrm{sc}=\int^\infty_{r} dr' n(r')\sigma(\vep_\nu) = \dot m\left(\frac{r_s}{r}\right)^{1/2}, 
\label{eq:tau_sc}
\end{equation}
where $n(r)$ is the nucleon number density ($n(r)=\dot M/[4\pi
  r^2m_pu(r)]$), $\sigma(\vep_\nu)$ is the scattering cross section
for neutrino energy $\vep_\nu$, and the dimensionless mass accretion
rate $\dot m=\dot M/\dot M_\mathrm{Edd,\nu}$, respectively.  Here,
$\dot M_\mathrm{Edd,\nu}$ is the Eddington accretion rate defined by
\begin{equation}
\dot M_\mathrm{Edd,\nu}\equiv \frac{L_\mathrm{Edd,\nu}}{c^2}=\frac{4\pi GM
  m_p}{\sigma(\vep_\nu)c},
\end{equation}
where $L_\mathrm{Edd,\nu}$ is the Eddington luminosity of neutrinos,
$M$ is the mass of the central object, $m_p$ is the proton mass, and
$G$ is the gravitational constant. Since $\dot m$ depends on the
neutrino energy, we introduce $\dot m_{kT}$, which represents $\dot m$
for $\vep_\nu=kT$ with $k$ and $T$ being Boltzmann constant and the
matter temperature.

For supercritical accretion into black holes ($\dot m\ge 1$), it is
evident from Eq. (\ref{eq:tau_sc}) that there should be regions in the
flow where the neutrinos propagate diffusively. For our problem, it is
convenient to use not $\tau_\mathrm{sc}$ but the effective optical
depth
\begin{equation}
  \tau\equiv\frac{3}{2}\frac{u(r)}{c}\tau_\mathrm{sc}(r)=\frac{3}{2}\dot m\frac{r_s}{r}=\frac{3}{2}\dot m_{kT}\left(\frac{\vep_\nu}{kT}\right)^2\frac{r_s}{r}.
  \label{eq:tau}
\end{equation}
This value will replace the radial coordinate $r$ in the radiative
transfer equation in \S\ref{subsec:radiative_transfer}. It should be
noted that $\tau$ depends on $\vep_\nu$ as well due to the energy
dependence of the scattering cross section. In addition, we introduce
the dimensionless neutrino energy as
\begin{equation}
x\equiv \frac{\vep_\nu}{kT}.
\end{equation}

\subsection{Radiative Transfer Equation}\label{subsec:radiative_transfer}

Now we proceed to write down and solve the radiative transfer
equation. For the kinetic equation, we start from equation (18) of
\cite{blan81} for the neutrino occupation number $f_\nu(\bf
r,\vep_\nu)$,
\begin{equation}
\begin{array}{ll}
  \displaystyle\pdel{f_\nu}{t}+\bu\cdot\nabla f_\nu =
  &
  \nabla\cdot\left(\displaystyle\frac{c}{3\kappa(\vep_\nu)}\nabla
  f_\nu\right)\\
  &
  +\displaystyle\frac{1}{3}(\nabla\cdot\bu)\vep_\nu\displaystyle\pdel{f_\nu}{\vep_\nu}
  +j(\bf r,\vep_\nu), 
  \end{array}
  \label{eq:kinetic}
\end{equation}
where $\kappa({\bf r})\equiv n({\bf r})\sigma(\vep_\nu)$ is the
inverse of the scattering mean free path and $j({\bf r},\vep_\nu)$ is
the emissivity.  Here, we neglect the recoil term, which affects the
neutrino spectrum only for the regime of $\vep_\nu\simgt m_p c^2\sim
1$ GeV (typical neutrino energy is $\sim$ 10 MeV).  Substituting the
inflow velocity $\bu=-u(r)\hat{\bf r}$, where $\hat{\bf r}$ is the
radial unit vector, and taking into account the spherical symmetry,
Eq. (\ref{eq:kinetic}) becomes
\begin{eqnarray}
  \pdel{f_\nu}{\tilde t}=\tau\pddel{f_\nu}{\tau}-\left(2\tau+\frac{3}{2}\right)\pdel{f_\nu}{\tau}-\frac{1}{2}x\pdel{f_\nu}{x}
  +\frac{\tau j}{3c\kappa(x)} ,
  \label{eq:kinetic_nodim}
\end{eqnarray}
where $\tilde t\equiv 3c\kappa t/\tau$ is the dimensionless time. The
dimensionless spectral energy flux $F(r,\vep_\nu)$ \citep{blan81} is
written in the new variables as
\begin{equation}
  F(\tau,x)\propto x^2\left(\frac{2\tau}{3\dot m_{kT}}\right)^{1/2}
  \left[\left(\frac{2\tau}{3}+1\right)\pdel{f_\nu}{\tau}+\frac{1}{3}x\pdel{f_\nu}{x}\right].
  \label{eq:flux_tau}
\end{equation}
Note that we are interested in the spectrum at a certain radius, not
the optical depth, which depends on both the radius and neutrino
energy. Thus, by combining Eqs. (\ref{eq:tau}) and
(\ref{eq:flux_tau}), we evaluate the spectral energy flux at a certain
radius as
\begin{equation}
  F(\tau,x)|_{r}\propto x^3
  \left[\left(\frac{2\tau}{3}+1\right)\left.\pdel{f_\nu}{\tau}\right|_r+\frac{1}{3}x\pdel{f_\nu}{x}\right].
  \label{eq:flux}
\end{equation}

\subsection{Analytic Solution}

In this subsection, we neglect the last term in the right hand side of
Eq.  (\ref{eq:kinetic_nodim}) in order to obtain an analytic solution.
This is because this term changes the number of neutrinos, which would
have minor contribution between the energy sphere and transport
sphere (see \S \ref{sec:neutrinosphere}).  Following \cite{payn81}, we solve Eq.
(\ref{eq:kinetic_nodim}) using variable separation with the form
\begin{equation}
  f_\nu(\tau,x)=R(\tau)\tau^{5/2}x^{-\alpha}.
\end{equation}
Here $\alpha$ becomes an eigenvalue of the following confluent
hypergeometric differential equation
\begin{equation}
  \tau\ddel{R}{\tau}+\left(\frac{7}{2}-2\tau\right)\del{R}{\tau}+\left(\frac{\alpha-10}{2}\right)R=0.
  \label{eq:hypergeometric}
\end{equation}
The physical solution of Eq. (\ref{eq:hypergeometric}) fulfills a
constant spectral flux of neutrinos as $\tau\to 0$ and adiabatic
compression of the neutrinos $\tau\to\infty$.  The relevant solution
can be evaluated as an infinite sum of generalized Laguerre
polynomials $L^{5/2}_n(2\tau)$. The corresponding eigenvalues
$\alpha_n$ are given by
\begin{equation}
  \alpha_n=4n+10;~~~~~~n=0,1,2,.....
\end{equation}
These values are different from \cite{payn81} because the cross
section depends on energy for the current case.  Note that this
spectral index does not imply the observable spectrum at a certain
radius because $\tau$ depends on not only a radius but also the
considered neutrino energy (see Eq. \ref{eq:tau}). In order to obtain
a spectral flux at $r$, we should multiply Eq. \eqref{eq:flux} by
$x^3$ because $\left.\partial f_\nu/\partial \tau\right|_r\propto
\tau^{3/2}\propto x^3 r^{-3/2}$ for $\tau\to 0$ (the other terms drop
faster than this term). Thus, the hardest spectral component (i.e.,
$n=0$) of $F$ becomes $\propto\vep_\nu^{-4}$.

In principle, the global solution of Eq. \eqref{eq:kinetic_nodim} can
be given by summing up infinite series expressed by $L^{5/2}_n(2\tau)
x^{-\alpha_n}$ with coefficients determined by the boundary
condition. In fact, \cite{payn81} gave the analytic solution with the
delta-function distribution function at the injection optical depth,
in which all coefficients are expressed by generalized Laguerre
Polynomials and Gamma functions (see Eqs. 10 and 11 in their
paper). However, the analytic expression for the arbitrary boundary
condition is not always representable using known functions. Thus, in
the following we solve Eq. \eqref{eq:kinetic_nodim} numerically with
the thermal distributions of neutrinos at the energy sphere as a
boundary condition.

\subsection{Numerical Solution}
\label{sec:numerical}

In this subsection, we present our numerical solution of
Eq. (\ref{eq:kinetic_nodim}). The last term is omitted again because the interested region is between the energy sphere and transport sphere (see \S \ref{sec:neutrinosphere}). We use the relaxation method for the
boundary problem \citep[e.g.,][]{pres92}, in which the stationary
solution is achieved by infinitely long exposure of the time dependent
equation.

In solving Eq. (\ref{eq:kinetic_nodim}), we change this equation to a
finite-difference form using
\begin{eqnarray}
&&\pddel{f_\nu}{\tau}=\frac{2}{\delta \tau_{i}+\delta \tau_{i+1}}\left(\frac{f_\nu^{i+1,j}-f_\nu^{i,j}}{\delta \tau_{i+1}}-\frac{f_\nu^{i,j}-f_\nu^{i-1,j}}{\delta \tau_{i}}\right), \\
&&\pdel{f_\nu}{\tau}=\frac{f_\nu^{i+1,j}-f_\nu^{i-1,j}}{\delta \tau_i+\delta \tau_{{i+1}}},\\
&&\frac{1}{2}x\pdel{f_\nu}{x}=\frac{1}{2}x_j\frac{f_\nu^{i,j+1}-f_\nu^{i,j-1}}{\delta x_j+\delta x_{{j+1}}},
\end{eqnarray}
where i and j denote the grid point of $\tau$ and $x$, respectively.
The grid points are determined by the rule as
\begin{eqnarray}
&&\tau_i=\tau_{i-1}+\delta \tau_i,\\
&&x_j=x_{j-1}+\delta x_j,\\
&&\delta \tau_i=r_\tau \delta\tau_{i-1},\\
&&\delta x_j=r_x \delta x_{j-1},
\end{eqnarray}
where $r_\tau$ and $r_x$ are constants larger than unity.  We set
$\delta \tau_{1}/\tau_{1}=\delta x_{1}/x_{1}=0.02$.  The calculations
are performed on a grid of 200 zones for $\tau$ from 0.01 up to
$\tau_0$ and 500 zones for $x$ from $0.1$ to $100$.  A test
calculation and comparison with an exact solution are given in
Appendix.

The boundary condition is given at $\tau_0$ as 
\begin{equation}
f_\nu(\tau_0,x)=\frac{\tau_0^{5/2}}{x^5}\frac{1}{1+e^x},
\end{equation}
where the factor $\tau_0^{5/2}/x^5$ means the correction, which leads
to the thermal distribution function at a radius $r$. In this study,
we set the above boundary condition with only one parameter $\tau_0$
for simplicity. In order to make more realistic boundary condition, we
should consider the microphysical processes that change the neutrino
number and energy in detail, which is beyond the scope of this paper.
Here we do not care about the normalization factor because all
equations solved in this study are linear to $f_\nu$ as we omit the
source term $j$. Needless to say, we should care about the
normalization with detailed source term because neutrino is fermion so
that there is a significant effect by Pauli blocking for $f\sim 1$.

\begin{figure}
  \centering
  \includegraphics[width=.5\textwidth]{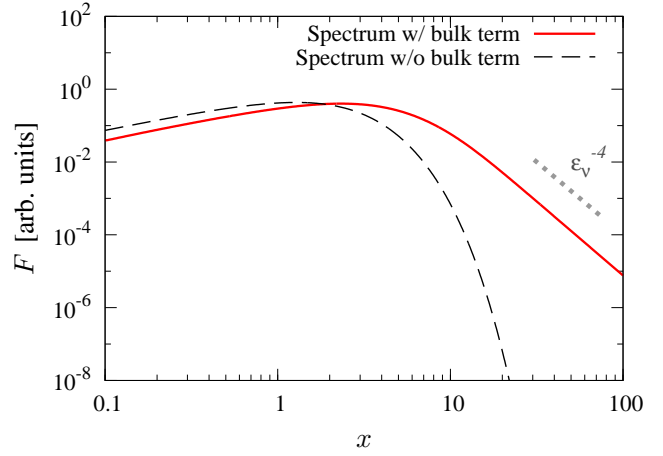}
  \caption{The emergent spectral energy flux estimated by the
    numerical solution. The boundary condition is given at
    $\tau_0=5$. The correction for the conversion from $\tau$-space to $r$-space is included (see text for details). The red solid line is the solution of the full
    equation of Eq. \eqref{eq:kinetic_nodim}, while the black dashed
    line is the solution of the kinetic equation without the bulk term
    (i.e., thermal spectrum). The grey dotted line represents the
    power-law spectrum of $\varepsilon_\nu^{-4}$, which is the
    analytic solution (see text for details).}
  \label{fig:numerical}
\end{figure}

In Figure \ref{fig:numerical}, we show the numerical solution of the
dimensionless spectral energy flux obtained by solving
Eqs. (\ref{eq:kinetic_nodim}) and (\ref{eq:flux}).  The red solid line
represents the emergent spectrum of full equation and the black thin
dashed line is spectrum obtained by the kinetic equation {\it without}
bulk term (i.e.,
$\frac{1}{3}(\nabla\cdot\bu)\varepsilon_\nu\pdel{f_\nu}{\varepsilon_\nu}$
in Eq. \ref{eq:kinetic}), that is, a thermal spectrum.  One can see
that neutrinos are upscattered by the infalling material and the
nonthermal spectrum is generated.  As indicated by grey-dotted line,
the emergent spectrum is power law with $\varepsilon_\nu^{-4}$ for
$x=\varepsilon_\nu/kT\simgt 10$, which is consistent with the analytic
solution with $n=0$ obtained in the previous section.  In this
calculation, the boundary condition is given at $\tau_0=5$, where
$f_\nu$ has a thermal distribution with a temperature, $T$.\footnote{
  According to numerical simulations of core-collapse supernovae,
  which include detailed microphysics and the radiative transfer, the
  temperature of $\nu_X$ ranges from 4 MeV to 10 MeV \citep[see][for a
    collective reference of recent numerical
    simulations]{hori09}. These values can be used in the case of long
  GRBs. As for short GRBs, \cite{seti06} showed that the average
  energy of $\nu_X$ range from $\sim 5$ MeV to $\sim 27$ MeV,
  corresponding to the temperature from $\sim 2$ MeV to $\sim 9$ MeV
  with vanishing chemical potential, similar to values of CCSNe.}  The
flux is estimated at $\tau=0.01$, where the spectral evolution is
almost completed.  The normalization of both spectra is determined by
the total number flux, $\int (F/x) dx$, being unity.

\begin{figure}
  \centering
  \includegraphics[width=.5\textwidth]{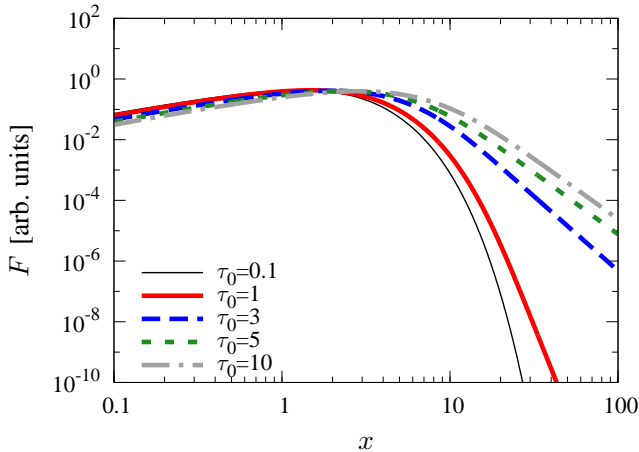}
  \caption{The emergent spectral energy flux with different $\tau_0$
    indicated by different lines. The higher $\tau_0$ leads to the
    harder spectrum due to efficient up-scattering by the infalling
    material.}
  \label{fig:tau}
\end{figure}

We show the different solutions with different $\tau_0$ in Figure
\ref{fig:tau}. It is obvious that higher $\tau_0$ leads to harder
spectrum. It should be noted that the position of the energy sphere
depends on the elementary process such as $\nu e\leftrightarrow\nu e$,
$e^+e^-\leftrightarrow\nu\bar\nu$, and $NN$ bremsstrahlung. The first
process is related to the thermalization and the others are related to
both the thermalization and emission/absorption.  The dominant
thermalization process depends on the background fluid temperature,
density, and abundance, which are much beyond the scope of this paper.  Thus, we simply parametrize the
injection $\tau_0$ and see the dependences on it \citep[see
  also][]{raff01}.

\section{Non-thermal Neutrinos and their Annihilation}\label{sec:annihilation}

Now we move on to estimate the neutrino annihilation rate, which
strongly depends on neutrino energy.  The energy deposition rate via
neutrino-annihilation ($\nu+\bar\nu\to e^{+}+e^{-}$) is given by
\citep{good87,seti06}
\begin{equation}
\dot E_{\nu\bar\nu}=\mathcal{C} F_{3,\nu}F_{3,\bar\nu}\left(\frac{\bracket{\vep_{\nu}^{2}}\bracket{\vep_{\bar\nu}}+\bracket{\vep_{\bar\nu}^{2}}\bracket{\vep_{\nu}}}{\bracket{\vep_{\nu}}\bracket{\vep_{\bar\nu}}}\right),
\end{equation}
where $F_{i,\nu}=\int f_\nu\vep_\nu^i \mathrm{d}\vep_\nu$,
$\bracket{\vep_\nu}=F_{3,\nu}/F_{2,\nu}$, and
$\bracket{\vep_\nu^2}=F_{4,\nu}/F_{2,\nu}$, respectively. The factor
$\mathcal{C}$ includes the weak interaction coefficients and the
information of the angular distribution of the neutrinos so that to
calculate this factor we should determine the geometry of the
neutrino-emitting source.  Since this factor is expected not to change
significantly by including the neutrino acceleration process, we
concentrate on the effect of the spectral change from here.  For
simplicity, we assume the spectrum of $\nu$ and $\bar\nu$ are
identical. Then, we get
\begin{equation}
\dot E_{\nu\bar\nu}\propto \frac{F_{3,\nu}^{2}\bracket{\vep_{\nu}^{2}}}{\bracket{\vep_{\nu}}}.
  \label{eq:q}
\end{equation}
We can evaluate the amplification of the neutrino annihilation rate by
the accelerated component of neutrino produced by the bulk motion of
background matter using $F_{3,\nu}$, $\bracket{\vep_\nu}$, and
$\bracket{\vep_\nu^2}$. By assuming that the neutrino number flux
($F_{2,\nu}$) does not change by including this effect, we get
$F_{3,\nu}\propto \bracket{\vep_{\nu}}$, then $\dot
E_{\nu\bar\nu}\propto \bracket{\vep_{\nu}}\bracket{\vep_{\nu}^{2}}$
Therefore, we can evaluate the amplification only by
$\bracket{\vep_\nu}$ and $\bracket{\vep_\nu^2}$.
 
Table \ref{tab} shows the integrated values of emergent spectrum. It
is obvious that both the mean energy $\bracket{\varepsilon_\nu}$ and
the mean-square energy $\bracket{\varepsilon_\nu^2}$ increase compared
to the thermals spectrum due to up-scattering by the infalling
materials.  As a result, the neutrino annihilation rate is
significantly amplified by the accelerated component.  In addition, we
show the convergence check with higher resolution in this table (see
last two lines). Due to much more expensive numerical cost, we just
calculate the model with $\tau_0=1$ and confirm the validity of the
lower resolution calculation.

\begin{table}
\centering
\caption{Properties of numerical solutions}
\begin{tabular}{cccccc}
\hline
$\tau_0$ & $N_\tau^{~\dag}$ & $N_x^{~\ddag}$ &  $\displaystyle\frac{\bracket{\vep_\nu}^\S}{\bracket{\vep_\nu}_{thermal}}$ & $\displaystyle\frac{\bracket{\vep^2_\nu}}{\bracket{\vep^2_\nu}_{thermal}}$& $\displaystyle\frac{\mathcal{A}^{~\P}}{\mathcal{A}_{thermal}}$\\
\hline
0.1 &  200 &  500 & 1.01 & 1.02 & 1.03  \\
0.2 &  200 &  500 & 1.03 & 1.05 & 1.08  \\
0.5 &  200 &  500 & 1.07 & 1.16 & 1.24  \\
1.0 &  200 &  500 & 1.16 & 1.37 & 1.59  \\
1.5 &  200 &  500 & 1.26 & 1.65 & 2.08  \\
2.0 &  200 &  500 & 1.37 & 1.99 & 2.73  \\
3.0 &  200 &  500 & 1.60 & 2.83 & 4.52  \\
5.0 &  200 &  500 & 1.95 & 4.49 & 8.77  \\
7.0 &  200 &  500 & 2.18 & 5.72 & 12.5  \\
10.0 &  200 &  500 & 2.43 & 7.12 & 17.3  \\
\hline
1.0 &  500 &  1000 & 1.16 & 1.36 & 1.57  \\
1.0 &  750 &  1500 & 1.16 & 1.37 & 1.59  \\
\hline
\label{tab}
\end{tabular}
\begin{flushleft}
$^\dag$ Numerical grids for $\tau$. \\
$^\ddag$ Numerical grids for $x$.\\
$^\S$ The average energy of emerged spectrum.\\
$^\P$ The amplification of the neutrino-annihilation rate
  (Eq. \ref{eq:q}) compared to the thermal distribution.
\end{flushleft}
\end{table}

\section{Summary and Discussion}
\label{sec:summary}

In this paper, we consider the spectral change of neutrinos induced by
the scattering of the infalling materials and amplification of the
annihilation rate, which is one of the well-discussed jet production
mechanism of GRBs. We solve the kinetic equation of neutrinos in
spherically symmetric background flow and find that neutrinos are
successfully accelerated and partly form nonthermal spectrum. We find
that the accelerated neutrinos can significantly enhance the
annihilation rate by a factor of $\sim 10$, depending on the injection
optical depth.

In this study, we tried to demonstrate the effect of the up-scattering
by the bulk motion of the material and just assumed the injection
optical depth with parametric manner. More realistic injection is
obtained by the insight of the energy sphere, whose position is
determined by the neutrino-electron inelastic scattering as follows.
The scattering opacity at energy sphere, where the inelastic
scattering with electrons freeze out, can be calculated by the ratio
of cross section\footnote{The total cross section of neutrino-electron
  inelastic scattering is \cite{burr02}
  \begin{equation}
    \sigma_\e\sim\frac{3}{8}\sigma_0\frac{\vep_\nu kT}{(m_\e c^2)^2},
    \nonumber
  \end{equation}
  where $\sigma_0\sim 1.7\times10^{-44}$ cm$^2$ is reference neutrino
  cross section, $m_\e$ is the electron mass, $T$ is the temperature
  of electrons. On the other hand, the total cross section of
  neutron-neutrino elastic scattering is
  \begin{equation}
    \sigma_\n\sim\frac{\sigma_0}{4}\left(\frac{\vep_\nu}{m_\e c^2}\right)^2.
    \nonumber
  \end{equation}
  The reason why we employ the cross section for neutrons is that due
  to the electron capture ($p+e^-\to n+\nu_\e$) the neutron fraction
  increases, whereas the proton fraction decreases inside the
  neutrinosphere. Therefore, neutrons are dominant target particle for
  propagating neutrinos. However, it should be noted that the total
  cross section of proton-neutrino scattering differs from neutron
  only $\sim$ 20\%.  },
\begin{equation}
  \frac{\tau_\mathrm{sc}(r_\mathrm{es})}{\tau_\mathrm{es}}\sim\frac{n_\mathrm{N}\sigma_\mathrm{n}}{n_\mathrm{e}\sigma_\e}\sim\frac{2\vep_\nu}{3Y_\e kT(r_\mathrm{es})}
\end{equation}
where $\tau_\mathrm{es}$ is the optical depth of electron inelastic
scattering, $r_\mathrm{es}$ is the radius of energy sphere,
$n_\mathrm{N}$ and $n_\mathrm{e}$ are the number density of nucleons
(neutrons and protons) and electrons, and
$Y_\e=n_\mathrm{e}/n_\mathrm{N}$ is the electron fraction,
respectively.  Since $\tau_\mathrm{es}$ is definitely $2/3$ at the
energy sphere,
\begin{equation}
  \tau_\mathrm{sc}(r_\mathrm{es})\sim\frac{4\vep_\nu}{9Y_\e kT(r_{\mathrm{es}})}.
\end{equation}
The typical temperature of neutrinospheres is $\sim 4$ MeV
\citep{jank01} so that $\tau_\mathrm{sc}$ is larger than $2/3$ for
neutrinos of the energy $\vep_\nu\simgt 0.6 (Y_\e/0.1)$ MeV. The
typical value of $Y_\e$ is $\sim 0.1$ at the region where the electron
capture is significant so that almost all the neutrinos are trapped by
the nucleon elastic scattering at the energy sphere.  Although the
injection optical depth in this study should be energy dependent as
shown above, we neglect this effect for simplicity.

Next, we discuss about the neutrino species. In this paper, we
consider the region between the energy sphere and the transport
sphere, i.e. the optical depth is larger than unity for the
neutrino-nucleon elastic scattering. Note that usually these neutrino
spheres are coincident for $\nu_\e$ and $\bar\nu_\e$ due to the
presence of the charged current for these neutrinos so that the
neutrino acceleration studied in this paper is possible only for
$\nu_\mu$, $\nu_\tau$ and their anti particles. However, for the case
of neutron number density being much larger than protons' one, the
charged current reaction of $\bar\nu_\e$ ($\bar\nu_{e}+p\to n+e^{+}$)
is negligible so that the reactions relevant to $\bar\nu_{e}$ become
similar to those of heavier leptonic neutrinos.
The transport opacity for the neutral current scattering processes are
given by \citep{jank01}
\begin{eqnarray}
    \label{eq:kappasc}
    \kappa_\mathrm{sc}\sim\frac{5\alpha^2+1}{24}\frac{\sigma_0\bracket{\epsilon_\nu^2}}{(m_\e c^2)^2}\frac{\rho}{m_\u}(Y_\n+Y_\p).
\end{eqnarray}
Here $m_\u\sim 1.66\times 10^{-24}$ g is the atomic mass unit,
$\alpha=-1.26$, and $Y_\n=n_\n/n_N$ and $Y_\p=n_\p/n_N$ are the number
fractions of free neutrons and protons, i.e., their particle densities
normalized to the number density of nucleons, respectively.
In cases of $\nu_\e$ and $\bar\nu_\e$ also the charged-current
absorption on neutrons and protons, respectively, need to be taken
into account due to their large cross sections. The absorption opacity
is \citep{jank01}
\begin{eqnarray}
    \label{eq:kappaa}
    \kappa_a \sim\frac{3\alpha^2+1}{4}\frac{\sigma_0\bracket{\epsilon_\nu^2}}{(m_\e c^2)^2}\frac{\rho}{m_\u}
    \left\{
        \begin{array}{c}
            Y_\n\\
            Y_\p
        \end{array}
    \right\}.
\end{eqnarray}
From Eqs. (\ref{eq:kappasc}) and (\ref{eq:kappaa}) the scattering
dominates the opacity for $\bar\nu_\e$ provided $Y_\p< 0.26$. In this
case, the transport sphere and number sphere \citep{jank95, raff01}
separate from each other for $\bar\nu_\e$. The time stationary
solutions of hyperaccreting flow imply that $Y_{e}$ can be as small as
$\sim 0.1$ \citep{kawa07}, similar to the case of core-collapse
supernova (just above the neutrino sphere, $Y_{e}\sim$ 0.05 --
0.1). Therefore, it is expected that the acceleration of
$\bar\nu_{\e}$ would naturally occur in the collapsar system.

Our finding suggests that the detectability of MeV neutrinos is also
enhanced because the expected detection number $\propto
F_{2,\nu}\bracket{\vep^2_\nu}$.  If this neutrino acceleration works
only for $\nu_{X}$, the neutrino oscillation would produce
$\bar\nu_{\e}$, which is main observable for walter \v{C}erenkov
detectors.  \cite{suwa09b} estimated the expected number from the
hyperaccreting accretion flow with thermal spectrum of $kT=3$ MeV and
argued that GRBs are observable for $\simlt$ a few Mpc by
Super-Kamiokande and several Mpc by Mton detector. The neutrino
acceleration process thought in this paper would push out the
detectable horizon of MeV neutrino farther.

At last, we comment on our assumptions in this study. 
Firstly, we employed the diffusion limit for whole region, which is
not valid for optically thin region, $\tau_\mathrm{sc}\simlt$ 1.  The
conclusion, however, does not change if we somehow include effects of
optically thinness, since the spectral evolution is determined by the
region $\tau_\mathrm{sc}\simgt$ 1.
Secondly, we dropped out the recoil term from the kinetic equation.
The average energy of neutrinos is not affected by this because the
recoil term changes the spectrum for $\simgt 1$ GeV which is much
higher than the typical energy of neutrino spectrum. Thus, if we
include the recoil term (that is much more complicated than terms
included in this study), the conclusion does not change very much.
Thirdly, we fixed the background matter flow as free fall. The energy
gain of neutrinos must come from the matter so that the back reaction
should be included when the total energy of neutrinos reaches as large
as the matter kinetic energy. To compare these two quantities, more
detailed source term is necessary. The final answer can be obtained by
solving neutrino radiation hydrodynamic equations in self-consistent
way, which is far beyond the scope of this simple study.
Fourthly, we omitted the relativistic effects, e.g., the Doppler shift
and gravitational redshift. In order to include these effects, we
should reformulate by covariant formulation of the radiative transfer
in self-consistent way. One can find such a formulation in
\cite{shib11}.
Finally, the spherical symmetry is also one of the largest assumption
in this study. This assumption is partially valid because NDAF
solution has large disk height due to the large contribution of gas
for pressure (otherwise if the disk is supported by the centrifugal
force, the disk height is almost negligible compared to the disk
radius). Considering the disk structure, neutrinos can escape from the
accretion flow to the vertical direction before the spectral evolution
completes. Whether this effect amplifies or suppresses the neutrino
annihilation is not trivial because there are two possible opposite
effects; neutrinos emitted at deep position, which are experienced
significant acceleration, can more easily escape than the spherically
symmetric configuration, while the acceleration might not complete due
to the earlier escape.
To give more concrete result, more detailed calculation is strongly
required.

\section*{Acknowledgments}

We thank K. Asano, K. Kiuchi, H. Okawa, Y. Sekiguchi, M. Shibata, and
T. Takiwaki for stimulating discussion and M. Suwa for
proofreading. This work has been supported in part by the
Grants-in-Aid for the Scientific Research from the Ministry of
Education, Science and Culture of Japan (No. 23840023) and HPCI
Strategic Program of Japanese MEXT.

\appendix

\section{Test calculation}

\begin{figure}
  \centering
  \includegraphics[width=.45\textwidth]{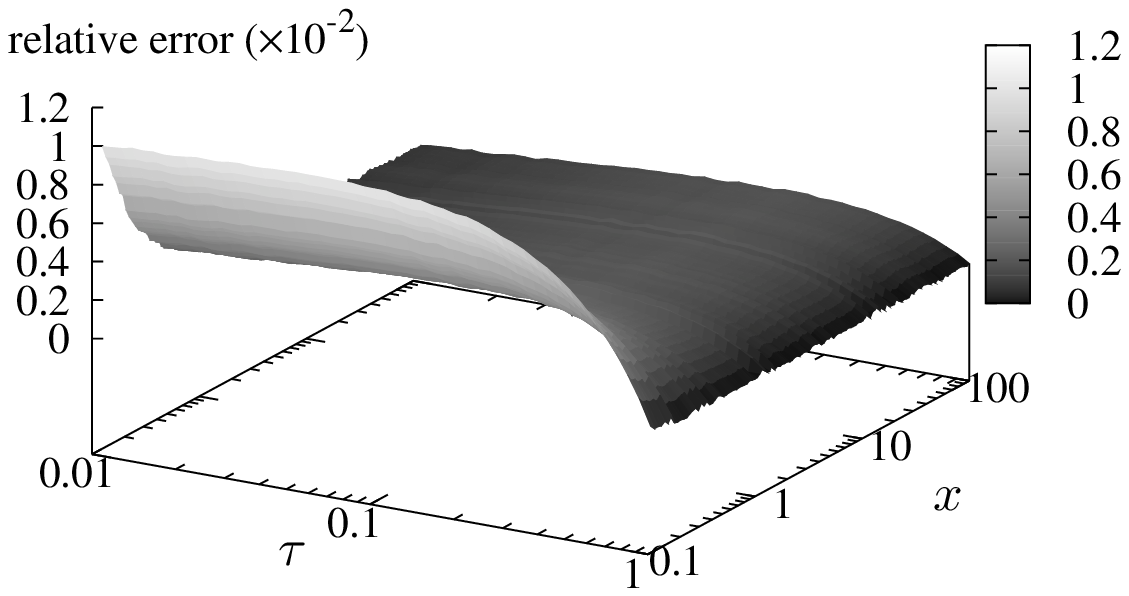}
  \caption{Absolute value of the relative error of the numerical
    solution in the case of $\tau_0=1$.}
  \label{fig:error}
\end{figure}

In this section, we show a test result of our numerical calculation. A
special solution of Eq. \eqref{eq:kinetic_nodim} is given by
\begin{eqnarray}
&&f_{\nu}(\tau,x)=\tau^{5/2}x^{-10},\label{eq:a1}
\end{eqnarray}
with $j=0$ and adequate boundary conditions. In Fig. \ref{fig:error},
we show the absolute value of the relative error, $|$1-(numerical
solution)$/$(exact solution)$|$ in the case where $\tau_{0}=1$ as an
example. The grid setup is the same as one used in
Sec. \ref{sec:numerical}. We find that the relative error is always
less than $\sim 10^{-2}$ for \eqref{eq:a1}. Hence, the error does not
affect the conclusion in \S \ref{sec:annihilation}.

\newcommand\aap{{A\&A}}%
\newcommand\apjl{{ApJ}}%
\newcommand\apj{{ApJ}}%
\newcommand\apjs{{ApJS}}%
\newcommand\mnras{{MNRAS}}%
\newcommand\prd{{Phys. Rev. D}}%
\newcommand\pasj{{PASJ}}%

\end{document}